\documentstyle[preprint,prl,aps]{revtex}

\begin{document}

\title{Effect of an External Field on Decoherence}

\author{R. F. O'Connell and Jian Zuo}

\address{Department of Physics and Astronomy, Louisiana State University
\\ Baton Rouge, Louisiana, 70803-4001, USA}

\date{\today}

\maketitle

\begin{abstract} "Decoherence of quantum superpositions through coupling to
engineered reservoirs" is the topic of a recent article by Myatt et al.
[Nature {\underline{403}}, 269 (2000)] which has attracted much interest
because of its relevance to current research in fundamental quantum
theory, quantum computation, teleportation, entanglement and the
quantum-classical interface.  However, the preponderance of theoretical
work on decoherence does not consider the effect of an
{\underline{external field}}.  Here, we present an analysis of such
an effect in the case of the random delta-correlated force discussed by
Myatt et al.

\end{abstract}

\newpage

"Decoherence of quantum superpositions through coupling to engineered
reservoirs"\cite{myatt} is the topic of a recent article which has
attracted much interest because of its relevance to current research in
fundamental quantum theory, quantum computation, teleportation,
entanglement and the quantum-classical interface.  As Schleich remarks in
an accompanying "News and Views" article, this is "-- a pioneering
experiment that engineers decoherence --"\cite{schleich}.  However, the
preponderance of theoretical work on decoherence\cite{giulini,ford1} does
not consider the effect of an {\underline{external field}}.  Here, we
present an analysis of such an effect in the case of the random
delta-correlated force discussed in Ref. 1.

Myatt et al.,\cite{myatt} used a linear Paul trap to confine single Be
ions in a harmonic potential and then prepared various superposition
states.  Next, they induced decoherence by coupling the single ion to a
reservoir which they controlled in various ways.  Such a reservoir gives
rise to an external force
$f(t)$ in the equation of motion of the system, in contrast to the usual
intrinsic fluctuation force $F(t)$ which arises from interaction with an
ambient thermal dissipative environment \cite{ford1}, which, of course,
will always be present, even at $T=0$.  Thus, the question arises as to
not only what is the dependence of the characteristic decoherence decay
time
$\tau_{d}$ on the separation $d$ of the superposition components, the
temperature
$T$ and the dissipative decay rate $\gamma$ (all of which come into play
when $f(t)=0$, the focus of most theoretical work) but also what is the
dependence on the parameters of the engineered reservoirs which give rise
to $f(t)$.  The existing experiments\cite{myatt} focused on the dependence
of $\tau_{d}$ on $d$ and demonstrated that $\tau_{d}\sim d^{-2}$.  This is
a familiar result predicted by the plethora of papers dealing with the
$f(t)=0$ situation but it does not give information on the dependence of
$\tau_{d}$ on the parameters of the externally-superimposed reservoir.
More details on the experimental results were given by Turchette et
al.\cite{turchette} and these authors also reviewed the theory of the
damping of a harmonic oscillator in a dissipative reservoir.  Whereas the
latter gives rise to a fluctuation force on the oscillator which is
related to the dissipation via the fluctuation-dissipation theorem, an
externally engineered situation requires an additional analysis, as is
made clear in \cite{ford3}.

Much of the discussion of decoherence\cite{giulini,ford1} has been in
terms of a particle moving in one dimension that is placed in an initial
superposition state (a Schr\"{o}dinger "cat" state) corresponding to two
widely separated Gaussian wave packets.  The corresponding wave function
has the form

\begin{equation}
\psi (x,0)=\frac{1}{(8\pi \sigma ^{2})^{1/4}(1+e^{-d^{2}/8\sigma
^{2}})^{1/2}}\left(\exp \left\{-\frac{(x-\frac{d}{2})^{2}}{4\sigma
^{2}}\right\}+\exp
\left\{-\frac{(x+\frac{d}{2})^{2}}{4\sigma ^{2}}\right\}\right),
\label{1efd}
\end{equation}
where $d$ is the separation and $\sigma^{2}$ is the variance of
each packet.  It is clear that the spatial probability distribution
$P(x,0)$ for this superposition of two states consists of the sum of the
probability distributions for the individual states plus an interference
term.  In the absence of dissipation \cite{ford4,ford}, one proceeds by
calculating $\psi (x,t)$ from which $P(x,t)$ readily follows.  However,
when dissipation is present it is necessary to use a density matrix
approach \cite{ford1} which, when combined with the use of quantum
probability functions, led to an expression for $P(x,t)$ of the form

\begin{equation} P(x,t)=P_{1}(x,t)+P_{2}(x,t)+P_{I}(x,t)\cos [f(t)],
\label{2efd}
\end{equation}
where $P_{1}$ and $P_{2}$ correspond to the time dependent
probability distributions for the separate wave packets and the third term
is an interference term.  The latter is characterized by a cosine factor
(which varies in time according to a known function $f(t)$\cite{ford1})
which is multiplied by an amplitude factor $P_{I}(x,t)$, which is found to
decay in time.  The disappearance of the interference term, that is the
decoherence, is measured by defining an attenuation coefficient
$a(t)$, which is the ratio of the factor multiplying the oscillatory term
to twice the geometric mean of the first two terms, i.e.

\begin{equation}
a(t)=\frac{P_{I}(x,t)}{2\left[ P_{1}(x,t)P_{2}(x,t)\right]^{1/2}}.
\label{3efd}
\end{equation} We should mention that, in the literature, one finds
various measures of decoherence, based on decay of diagonal and
off-diagonal density matrix elements or probability distributions in phase
space, momentum space or coordinate space\cite{murakami} but we consider
the latter to be the most desirable because it is closest to experiment.
Thus, returning to (\ref{3efd}), what we have found\cite{ford1} is that
$a(t)$ depends crucially on the spreading of the wave packets
corresponding to the individual states.

For $f(t)=0$, this spreading
arises from the possible intrinsic spreading associated with the uncertainty
principle, thermal spreading and spreading due to dissipative ($\gamma$)
effects.  Explicitly, for a free particle described by a single wave packet, the
width
after a time
$t$ is
$w(t)$, given by\cite{ford1}

\begin{equation}
w^{2}(t)=\sigma^{2}-\frac{\left[x(t_{1}),x(t_{1}+t)\right]^{2}}{4\sigma^{2}}
+s_{0}(t), \label{4efd}
\end{equation}
where $\sigma$ is the initial width and $s_{0}(t)$ is the mean
square displacement (discussed in more detail below).  For the attenuation
coefficient in the case of a {\underline{free}} particle we have the
formula\cite{ford1}

\begin{equation}
a(t)=\exp\left\{-\frac{s_{0}(t)d^{2}}{8\sigma^{2}w^{2}(t)}\right\}.
\label{5efd}
\end{equation}
In addition, the characteristic time for decay to occur, $\tau_{d}$ say, is
defined as usual
\cite{ford1,ford4,ford} as the time at which $a(t)=\exp(-1)$.

We now turn to the case where $f(t)\neq 0$ and we generalize from the case
of a free particle to that of an {\underline{oscillator}} potential, corresponding
to
the experiment described in \cite{myatt}.  For $f(t)\neq 0$, there is an
additional spreading of the wave packets, which we will now calculate.
Afterwards, we will turn to the role it plays in the calculation of $a(t)$.

Let $x(t)$ be the dynamical variable corresponding to the coordinate of
the wave function of the superposition state of the oscillator of Myatt et
al.\cite{myatt}.  As shown in Ref. \cite{ford2}, in the presence of an external
force
$f(t)$ in addition to the fluctuation force
$F(t)$, the steady-state motion can be described by means of a generalized
quantum Langevin equation

\begin{equation}
m\ddot{x}+\int^{t}_{-\infty}dt^{\prime}\mu(t-t^{\prime})\dot{x}(t^{\prime})+Kx=
F(t)+f(t), \label{6efd}
\end{equation} where $\mu(t)$ is the memory function, $K$ is the
oscillator force constant $(K=m\omega^{2}_{0})$, where $\omega_{0}$ is the
oscillator
frequency, and
$F(t)$ is a fluctuating operator force with mean $\langle F(t)\rangle =0$.  The
steady-state solution of (\ref{6efd}) can be written as

\begin{eqnarray} x(t) &=&
\int^{t}_{-\infty}dt^{\prime}G(t-t^{\prime})[F(t^{\prime} )+f(t^{\prime})]
\nonumber \\
&\equiv& x_{s}(t)+x_{d}(t), \label{7efd}
\end{eqnarray} where $x_{s}(t)$ is the stationery solution and $x_{d}$ is
due to the driven motion.  Also, $G(t)$, the Green function, is given by

\begin{equation}
G(t)=\frac{1}{2\pi}\int^{\infty}_{-\infty}d\omega\alpha(\omega +i0^{+}
)e^{-i\omega t}, \label{8efd}
\end{equation} with $\alpha (z)$ the familiar response function

\begin{equation}
\alpha(z)=\frac{1}{-mz^{2}-iz\tilde{\mu}(z)+K}. \label{9efd}
\end{equation} In addition

\begin{eqnarray}
\tilde{\mu}(z) &=& \int^{\infty}_{0}dt\mu (t)e^{izt} \nonumber \\ &\equiv&
m\gamma (z), \label{10efd}
\end{eqnarray} is the Fourier transform of the memory function and it
characterizes the reservoir \cite{ford1,ford2}.  The fact that $\alpha(z)$ does
not depend on $f(t)$ follows simply by taking the Fourier transform of
(\ref{7efd}) which enables the solution to be written in Fourier transform
language as

\begin{equation}
\tilde{x}(\omega )=\alpha(\omega )[\tilde{F}(\omega )+\tilde{f}(\omega )],
\label{11efd}
\end{equation} where superposed tildes indicate Fourier transforms.

Because of the linearity of the oscillator, it is clear that the motion of
the driven oscillator will be a superposition of a driven mean motion and
a motion about the mean that is identical with the motion about the
equilibrium state \cite{ford3}.  The starting-point of our calculation is the
correlation

\begin{eqnarray}
\frac{1}{2}\langle x &(& t)x(t^{\prime})+x(t^{\prime})x(t)\rangle \equiv
C(t-t^{\prime})\equiv C_{0}+C_{d} \nonumber \\ &&
{}=\frac{\hbar}{\pi}\int^{\infty}_{0}d\omega
{\textnormal{Im}}\{\alpha(\omega +i0^{+})\}\coth\frac{\hbar\omega}{2kT}
\cos\omega(t-t^{\prime})+C_{d}, \label{12efd}
\end{eqnarray} where $C_{0}$ and $C_{d}$ are the contribution due to
$F(t)$ and $f(t)$, respectively.  It follows that the mean-square
displacement (which characterizes the spreading of the wave packet) is

\begin{eqnarray} s(t) &\equiv& \langle [x(t)-x(0)]^{2}\rangle
=2\{C(0)-C(t)\}
\nonumber \\ &=& \frac{2\hbar}{\pi}\int^{\infty}_{0}
d\omega{\textnormal{Im}}\{\alpha
(\omega+i0^{+})\}\coth\frac{\hbar\omega}{2kT}(1-\cos\omega t)+s_{d},
\label{13efd}
\end{eqnarray} where $C(t)$ is given by (\ref{12efd}) and $s_{d}$ is the
contribution due to the "driven motion".

Here, we have used the fact that since $<F(t)>=0$ and since there is no
correlation between $F(t)$ and $f(t)$, it is clear that

\begin{equation} s(t)=s_{0}(t)+s_{d}(t), \label{14efd}
\end{equation} where $s_{0}$ denotes the contributions due to $F(t)$.
Since
$s_{0}(t)$ has been calculated in detail, in Ref. \cite{ford1}, which
considers entanglement between the system and the environment at the
initial time $t=0$, we will henceforth concentrate on
$s_{d}$.  Consider that the external force is applied at $t=0$.  It
follows from (\ref{7efd}) that

\begin{equation}
x_{d}(t)=\int^{t}_{0}dt^{\prime}G(t-t^{\prime})f(t^{\prime}). \label{15efd}
\end{equation} Since $x_{d}(0)=0$, it follows that

\begin{eqnarray} s_{d}(t) &=& \langle x^{2}_{d}(t)\rangle \nonumber \\ &=&
\int^{t}_{0}dt^{\prime}\int^{t}_{0}dt^{\prime\prime}G(t-t^{\prime})
G(t-t^{\prime\prime})g(t^{\prime}-t^{\prime\prime}), \label{16efd}
\end{eqnarray} where

\begin{equation} g(t^{\prime}-t^{\prime\prime})=\langle
f(t^{\prime})f(t^{\prime\prime})\rangle. \label{17efd}
\end{equation}

Further progress clearly depends on the nature of $f(t)$ but, keeping in
mind the existing experiments
\cite{myatt,turchette}, let us consider a random delta-correlated force so
that

\begin{equation} g(t^{\prime}-t^{\prime\prime})=g\delta
(t^{\prime}-t^{\prime\prime}),
\label{18efd}
\end{equation} where $g$ is time-independent. Hence, substituting
(\ref{18efd}) in (\ref{16efd}), we obtain

\begin{equation} s_{d}=g\int^{t}_{0}dt^{\prime}G^{2}(t^{\prime}).
\label{19efd}
\end{equation}
In the case of the oscillator potential of Myatt et al. \cite{myatt}, we
find that in the case of Ohmic coupling ($\gamma (\omega )=\gamma =$
constant)

\begin{equation}
G(t)=e^{-(\gamma t/2)}~\frac{\sin\omega_{1}t}{m\omega_{1}}, \label{20efd}
\end{equation}
where

\begin{equation}
\omega^{2}_{1}=\omega^{2}_{0}-(\gamma /2)^{2}. \label{21efd}
\end{equation}
Thus, substituting
(\ref{20efd}) in (\ref{19efd}), it follows that

\begin{equation}
s_{d}=\frac{g}{4m^{2}\gamma\omega^{_2}_{0}\omega^{2}_{1}}\left\{(1-e^{-\gamma
t})2\omega^{2}_{1}-e^{-\gamma
t}(\gamma^{2}\sin^{2}\omega_{1}t+\gamma\omega_{1}\sin2\omega_{1}t)\right\}.
\label{22efd}
\end{equation}
In the absence of dissipation $(\gamma\rightarrow 0)$ (which approximates the
experiment of Myatt et al.), (\ref{21efd}) and (\ref{22efd}) give

\begin{equation}
s_{d}\rightarrow\frac{g}{2m^{2}\omega^{2}_{0}}t\left\{1-\frac{\sin2\omega_{0}t}
{2\omega_{0}t}\right\}. \label{23efd}
\end{equation}

Again, for $\gamma\rightarrow 0$ and $T\rightarrow 0$ (absence of
dissipation and for negligibly low temperatures), it readily follows that

\begin{equation}
a(t)=\exp\left\{-\frac{s_{d}(t)d^{2}}{8\sigma^{2}[\sigma^{2}+s_{d}(t)]}\right\},
\label{24efd}
\end{equation}
where $\sigma$ is the initial width of the individual wave
packets.  Thus, the dependence on $d^{2}$ in the
numerator always emerges, regardless of the value of $s_{d}$.  We also note
the absence of a term analogous to the second term in (\ref{4efd}),
corresponding to the fact that, when $f(t)=0$, the width of the oscillator
wave function is constant in time whereas that of the free particle
continually increases.

It is clear from (\ref{24efd}) that the relative magnitudes of $s_{d}$ and
the initial variance $\sigma^{2}$ play a crucial role.  In particular,

\begin{equation}
a(t)\approx\exp\left\{-\frac{d^{2}}{8\sigma^{2}}\right\}~~~
{\textnormal{if}}~~s_{d}>>\sigma^{2}, \label{25efd}
\end{equation}
and

\begin{equation}
a(t)\approx\exp\left\{-\frac{s_{d}}{\sigma^{2}}~\frac{d^{2}}{8\sigma^{2}}
\right\}~~~{\textnormal{if}}~~s_{d}<<\sigma^{2}. \label{26efd}
\end{equation}
Thus, in the former case, the result for $a(t)$ is independent of $s_{d}$
i.e. independent of the external force $f(t)$.  In the latter case, using
(\ref{23efd}),  we see that

\begin{equation}
a(t)=\exp\left\{-\frac{t}{\tau_{0}}\left(1-\frac{\sin
2\omega_{0}t}{2\omega_{0}t}\right)\right\}, \label{27efd}
\end{equation}
where

\begin{equation}
\tau_{0}=\frac{16\sigma^{4}m^{2}\omega^{2}_{0}}{d^{2}g}. \label{28efd}
\end{equation}
For small times ($2\omega_{0}t<<1$) after the initial time $t=0$, we see that

\begin{eqnarray}
a(t) &=& \exp\left\{-\frac{t}{\tau_{0}}~\frac{(2\omega_{0}t)^{2}}{6}\right\}
\nonumber \\
&=&
\exp\left\{-\frac{gd^{2}}{24m^{2}\sigma^{4}}t^{3}\right\},~~~{\textnormal{if}}
~~~\omega_{0}t<<1, \label{29efd}
\end{eqnarray}
in which case the decay rate of decoherence is independent of $\omega_{0}$,
corresponding to free particle behaviour.  However, when $t$ further increases
there is
a change in the time behaviour until at the end of the first cycle at
$2\omega_{0}t=2\pi$, we see from (\ref{27efd}) that

\begin{equation}
a(t)=\exp\left\{-\frac{t}{\tau_{0}}\right\}. \label{30efd}
\end{equation}
In fact, as we go into the next and subsequent cycles, the
$\sin(2\omega_{0}t)/2\omega_{0}t$ term becomes more and more negligible so
that (\ref{30efd}) becomes more and more accurate as we go beyond the first
cycle.

It should also be noted that
(\ref{18efd}) also corresponds to a white-noise spectrum. However, it is
very different in nature than the white-noise spectrum associated with the
fluctuation force $F(t)$.  A random c-number field feeds energy into the
quantum particle (and, in fact, for a particle with negligibly weak coupling
to a heat bath and for either a zero or oscillator potential, it may be shown
that the energy of the particle increases linearly in time).  On the other
hand, in the case of a fluctuation force, we are necessarily dealing with a
heat bath; in other words, we have a dynamical system in which the particle
also loses energy due to the emission of bath excitations.  Thus, for
example, in the case where the white-noise spectrum is associated with an
equilibrium temperature
\cite{ford2}

\begin{equation}
\langle F(t^{\prime})F(t^{\prime\prime})\rangle =2m\gamma T\delta
(t^{\prime}-t^{\prime\prime}). \label{31efd}
\end{equation}
for the case of constant $\gamma$ and in the classical limit.  Moreover, the rate
of
work being done by the fluctuation force,
$P_{F}$ say, is given by \cite{li}

\begin{equation}
P_{F}=kT\gamma . \label{32efd}
\end{equation}
Thus, the rate of work being done by the fluctuation force is proportional
to the dissipation.  This is a manifestation of the general principle that,
at equilibrium, the energy lost by a particle due to dissipation is
compensated by the energy received from the fluctuation force.  Thus, there is a
crucial
difference between the effects of $f(t)$ and $F(t)$ so that, in particular, an
external
field that has a white noise spectrum can not be approximated by a weakly-coupled
thermal reservoir and, as a result, one must use the analysis given above.

Finally, it is clear that in order to
explore the larger parameter space (such as dependence on
$T,
\gamma$ and various choices of
$f(t)$ as well as on the potential), both further experiments and theoretical
work will be needed.  Some recent work has made inroads into this multi-
dimensional
parameter space.  First, for the problem considered above, we find that a non-
random
external force does not cause decoherence.  Second, in the absence of an external
field,
the Schr\"{o}dinger cat superposition has been examined for the case of an
oscillator
potential and high temperature \cite{ford5} and for the case of a free particle
subject
to the effects of the zero-point oscillations of the electromagnetic field
\cite{ford6}.

We are pleased to thank Professor G. W. Ford for many enlightening discussions.


\begin{references}
\bibitem{myatt} C. J. Myatt, B. E. King, Q. A. Turchette, C. A. Sackett,
D. Kielpinski, W. M. Itano, C. Monroe and D. J. Wineland, {\emph{Nature}}
{\bf{403}}, 269 (2000).

\bibitem{schleich} W. P. Schleich, {\emph{Nature}} {\bf{403}}, 256 (2000).

\bibitem{giulini} D. Giulini, E. Joos, C. Kiefer, J. Kupisch, I.-O.
Stamatescu, and H. D. Zeh, "Decoherence and the Appearance of a Classical
World in Quantum Theory" (Springer, New York, 1996).

\bibitem{ford1} G. W. Ford, J. T. Lewis, and R. F. O'Connell, {\emph{Phys.
Rev. A}} {\bf{64}}, 032101 (2001).

\bibitem{turchette} O. A. Turchette, C. J. Myatt, B. E. King, C. A.
Sackett, D. Kielpinski, W. M. Itano, C. Monroe, and D. J. Wineland,
{\emph{Phys. Rev. A}} {\bf{62}}, 053807 (2000).

\bibitem{ford3} G. W. Ford, and R. F. O'Connell {\emph{Opt. Commun.}}
{\bf{179}}, 451 (2000).  See especially Sec. 4.

\bibitem{ford4} G. W. Ford and R. F. O'Connell, Phys. Lett. A {\bf{286}},
87 (2001).

\bibitem{ford}
G. W. Ford and R. F. O'Connell, Am. J. Phys. {\bf{70}}, 319 (2002).

\bibitem{murakami} M. Murakami, G. W. Ford and R. F. O'Connell, Laser
Physics {\bf{13}}, 180 (2003).

\bibitem{ford2} G. W. Ford, J. T. Lewis, and R. F. O'Connell, {\emph{Phys.
Rev. A}} {\bf{37}}, 4419 (1988).

\bibitem{li} X. L. Li, G. W. Ford and R. F. O'Connell, Phys. Rev. E
{\bf{48}}, 1547 (1993).

\bibitem{ford5} G. W. Ford and R. F. O'Connell, Proceedings of
the Wigner Centennial Conference, Pecs, Hungary, 2002, [Acta
Phys. Hung. (to be published).

\bibitem{ford6} G. W. Ford and R. F. O'Connell, J.
Optics B {\bf{5}}, S609 (2003). Special Issue on Quantum Computing.
\end{references}
\end{document}